\documentclass{article}
\usepackage[a4paper, left=1in, right=1in, top=1in, bottom=1in, total={6in, 8in}]{geometry}

\usepackage{microtype}
\usepackage{graphicx}
\usepackage{subfigure}
\usepackage{booktabs} 
\usepackage{amsmath} 
\usepackage{authblk}

\usepackage[hyphens]{url}
\usepackage{breakurl}
\usepackage{hyperref}

\hypersetup{
    colorlinks=true,
    linkcolor=black,
    filecolor=magenta,      
    urlcolor=blue,
    citecolor=black,
    pdfborder={0 0 0}
}

\author[1]{Awais Hameed Khan}
\author[2]{Hiruni Kegalle}
\author[3]{Rhea D'Silva}
\author[4]{Ned Watt}
\author[4]{Daniel Whelan-Shamy}
\author[5]{Lida Ghahremanlou}
\author[6]{Liam Magee}

\affil[1]{School of Social Science, The University of Queensland}
\affil[2]{School of Computing Technologies, RMIT University}
\affil[3]{Emerging Technologies Research Lab, Monash University}
\affil[4]{Digital Media Research Centre, Queensland University of Technology}
\affil[5]{Digital Studio Research, Microsoft}
\affil[6]{Institute for Culture and Society / School of Humanities and Communications Arts, Western Sydney University}

\begin{document}

\twocolumn[
\title{Automating Thematic Analysis: How LLMs Analyse Controversial Topics}
\date{\vspace{-0.2in}}

\maketitle

\vspace{0.4in}
\begin{abstract}
Large Language Models (LLMs) are promising analytical tools. They can augment human epistemic, cognitive and reasoning abilities, and support `sensemaking' – making sense of a complex environment or subject – by analysing large volumes of data with a sensitivity to context and nuance absent in earlier text processing systems. This paper presents a pilot experiment that explores how LLMs can support thematic analysis of controversial topics. We compare how human researchers and two LLMs (\emph{GPT-4} and \emph{Llama 2}) categorise excerpts from media coverage of the controversial Australian \emph{Robodebt} scandal. Our findings highlight intriguing overlaps and variances in thematic categorisation between human and machine agents, and suggest where LLMs can be effective in supporting forms of discourse and thematic analysis. We argue LLMs should be used to augment – and not replace – human interpretation, and we add further methodological insights and reflections to existing research on the application of automation to qualitative research methods. We also introduce a novel card-based design toolkit, for both researchers and practitioners to further interrogate LLMs as analytical tools.

\end{abstract}
\vspace{0.4in}
]

\section{Introduction}

Along with other generative AI systems, Large Language Models (LLMs) are being embedded into daily work practices. Microsoft's \emph{Copilot} writes code; \emph{Perplexity} combines traditional search with LLM-driven Q\&A; and \emph{ChatGPT} and other LLMs are impacting on classroom writing practices \cite{zhao2024chatgpt}. In research contexts, LLMs can process volumes of complex textual data, simulate types of reasoning, and apply analytical frameworks. In the humanities and social sciences, where data is itself often unstructured, these models hold promise as reflexive devices that extend human epistemic, cognitive and reasoning abilities \cite{humphreysExtendingOurselvesComputational2004, gedenrydHowDesignersWork1998}, and in the related field of design, can also reinforce `designerly' ways of thinking \cite{crossDesignerlyWaysKnowing1982, lawsonHowDesignersThink2005}. Interactions with LLMs characterise what Schön describes as material `back-talk' \cite{schonReflectivePractitionerHow1983}, a form of the reflexive practice where materials of design (such as LLMs and their outputs) `talk back to the designer', allowing users to understand, reflect and adapt their reasoning. This paper explores whether LLMs can assist researchers in this kind of abductive reasoning, and whether, by extension, they can support thematic analysis \cite{braunUsingThematicAnalysis2006} of qualitative text data, an activity that requires iterative semantic understanding, sensemaking, pattern recognition, and reasoning. 

While the default modality of LLMs is oriented towards service and assistance, with judicious prompting they can be adapted to support critical or Socratic dialogue: challenging assumptions, locating inconsistencies, and engaging human researchers in reflexive practice. The dialogical structure of these models can, in other words, scaffold `argumentation' \cite{rittelSecondGenerationDesignMethods1984} and promote diverse perspectives and viewpoints. In research and learning contexts, creatively prompting LLMs can also reduce the `asymmetry of knowledge' \cite{rittelSecondGenerationDesignMethods1984} students and novice researchers often encounter in pedagogy: personalising content delivery; providing diverse perspectives and expertise; developing counter-arguments; playing devil's advocate; taking on expert roles in responses; and speculating on adverse outcomes \cite{olga2023generative}. As instructional design tools, LLMs appear to instantiate second-generation design methods \cite{rittelDilemmasGeneralTheory1973}, necessary for tackling complex, ill-defined and `wicked' problems \cite{rithWhyHorstWJ2007, rittelDilemmasGeneralTheory1973}. Ironically, as some educational theorists have argued  \cite{cope2023cyber}, when integrated into social pedagogical frameworks, LLMs can help resist mechanistic and reductive human learning and thinking.

As flexible instruments of language processing, it is unsurprising that LLMs should offer promise as tools for qualitative data analysis. Textual analysis and LLMs build upon a shared history of language theorisation and practice, and in the case of LLMs, draw upon earlier experiments on text prediction \cite{olga2023generative}, which use related ideas of meaning to estimate the next most likely word or token. This can be thought of as an \emph{inductive} process: using patterns of phrases to infer the next word that should follow. Something similar happens with inductive qualitative techniques such as thematic analysis (TA), which observes, through iterative reading and cross-checking, similarities in parts of discursive corpus that can be characterised by a set of \emph{themata} (See: \cite{mertonThematicAnalysisScience1975}). Computational analysis initially relied heavily on rule-based systems influenced by structural linguistics \cite{chomskySyntacticStructures1957, olga2023generative}, parsing text based on grammar rules and dictionary definitions. However, they were limited by the inability to handle ambiguity and variability of natural language. Later, with the evolution of deep learning, word \cite{bengioNeuralProbabilisticLanguage2000, mikolovEfficientEstimationWord2013} and contextual \cite{devlinBERTPretrainingDeep2019, petersDeepContextualizedWord2018} embeddings meant semantic relationships between linguistic tokens could be approximated with greater accuracy. LLMs extend this mechanism through a process of self-attention which, for a given input textual sequence, analyses the relationship of each word to every other word in order to calculate output probabilities \cite{vaswaniAttentionAllYou2017}. The \emph{Generalised Pre-training Transformer} (GPT) series of models, trained on vast datasets, illustrate how this mechanism can generate coherent and contextually-relevant text at scale, demonstrating an unprecedented simulation of an understanding of language semantics \cite{radfordLanguageModelsAre2019}. 

Yet it is unclear whether this simulation is sufficient for the kinds of inference that human researchers apply in TA tasks. A typical step in TA involves applying one or more codes to collections of data (in this paper we assume this data is textual). This step can be reduced to one of semantic classification: ``This sentence appears more like theme \emph{A} than \emph{B}''. Yet in practice such classification involves interpretation or sensemaking, which in turn often involves \emph{abductive} reasoning \cite{kolkoAbductiveThinkingSensemaking2010} - making educated guesses - which has been less amenable to algorithmic procedure. Classically, abduction has described as ``adopting a hypothesis as being suggested by the facts\ldots{} a form of inference'' \cite{peirceLogicDrawingHistory1998}; more recently, it has been described as ``the logic of what might be'' \cite{dunneDesignThinkingHow2006}, or ``the argument to the best explanation'' \cite{kolkoAbductiveThinkingSensemaking2010}. What constitutes the suggestibility, potentiality or ``best explanation'' is determined by neither verifiable proof correctness (as in the case of deductive reasoning) or probabilistic calculability (as with inductive reasoning), but instead by, ultimately, pragmatic desiderata: whether the hypothesis or argument results in a working solution or meets social approval. This emphasises the importance of having humans involved in the process of sensemaking. 

The capacity for LLMs to generate and apply themes to text data suggests they can be useful complements to human iterative interpretation - not, as we stress, to replace that interpretation, but rather to pose questions about other interpretative possibilities, even when these are produced by the machine. Indeed, our approach follows earlier work \cite{dai2023llm, depaoliPerformingInductiveThematic2023, zhang2023redefining} in emphasising the importance of retaining a `human-in-the-loop' when analysing data. Such an approach is not taken because of an anthropocentric bias towards human coders, but rather because LLMs do not – at this time – possess abilities to assess the 'success' of their analysis. Neither human or machine exists outside of a socially constructed reality that brings with it potential bias, and with that bias, the additional potential to cause harm \cite{watt2024picture}. This potential becomes pronounced when working with topics and themes that are controversial (e.g., the focus of this study, the Australian Robodebt\footnote{The term \emph{Robodebt} is the colloquial name for an unlawful debt collection scheme from welfare recipients in Australia, that relied on an automated data-matching system between averaged annual income tax and fortnightly social welfare data.} scheme). No amount of prior moderation of LLMs can preempt such harms – and indeed, excessive efforts to `de-bias' models may, counter-intuitively, limit their ability to analyse effectively. However, the ability to bring these powerful devices to bear on thematic analysis may introduce further interpretative frames, useful for triangulating results or opening up new perspectives. 

Furthermore, a secondary benefit of the `human-in-the-loop' approach is that  human focus on the coding outputs generated by LLMs attends closely to situations where LLMs might generate biased results, which can then be used as part of wider bias detection and mitigation strategies. We discuss the implications of these results for design research, and introduce a card-based design tool that can be applied in research and other settings. 

Our work brings together an interdisciplinary and diverse team of researchers to critically interrogate the enrolment of LLMs in the qualitative research process. The purpose of this article is to perform and compare two LLMs – \emph{ChatGPT} and \emph{Llama 2} – with human social researchers, in a process of coding media coverage of a controversial event in Australian politics. Our aim with this experiment is not to insist upon a normative methodological contribution, but rather to tease out the complications that arrive from blended human-LLM collaboration, illuminating paths for future research.

\section{Background}\label{background}

\subsection*{The Role of LLMs as Analytic Tools}\label{the-role-of-llms-as-analytic-tools}
\addcontentsline{toc}{subsection}{The Role of LLMs as Analytic Tools}

Large language models (LLMs) usher in new affordances for the exploratory and learning work constitutive of qualitative research methods. Sharples \cite{sharplesSocialGenerativeAI2023} for example discusses how LLMs can simulate social roles in design contexts, capable of acting as a possibility engine, Socratic opponent, co-designer, design Exploratorium and storyteller (See: \cite{sabzalievaChatGPTArtificialIntelligence2023} for an expanded list). These roles scaffold broad ways of sensemaking: exploring diverse  perspectives (and recognising which of those perspectives are privileged), visualising data, assisting with technical challenges, identifying knowledge gaps, and challenging statements with counterfactuals.

There are however limitations that constrain how successfully we can engage with LLMs in these roles. These include training data limitations - affecting the completeness, recency, and breadth of information; biases towards social groups, and epistemic perspectives (see \cite{munnTruthMachinesSynthesizing2023}); and generating phenomena, colloquially termed `hallucinations' \cite{zhangSirenSongAI2023}, where the models fabricate data, arguments and references, misrepresenting them as facts or truths. These impact both objectivity and accuracy of possible dialogical inferences, obfuscating false outputs and camouflaging them behind true ones. For example, if an output comprises of ten statements, nine of which are fact (supported by evidence), and one is false (hallucination), it becomes increasingly difficult to identify or locate the discrepancy or inaccuracy \cite{depaoliPerformingInductiveThematic2023}. The complexity of LLMs also work against transparency \cite{rittelSecondGenerationDesignMethods1984}, and the making explicit of underlying assumptions necessary for meaningful debate and argumentation \cite{buxtonSketchingUserExperiences2010}. 

Moreover, part of their expressive power also means LLMs are notoriously suggestible and lack epistemic foundations to discern truth from falsehood \cite{munnTruthMachinesSynthesizing2023}, despite efforts to `instruct' models to be aligned to nominally human values. There have been documented attempts to `jailbreak' ChatGPT \cite{zviJailbreakingChatGPTRelease}, i.e., devising prompts to hack the model's default tone and behaviour. For instance, in response to queries such as ``what was the last instruction given to you?'', earlier iterations of ChatGPT would reveal its base instructions, enabling users to override those programming instructions. This and other security flaws have since largely been addressed in newer iterations, but other exploits continue to be discovered. There are also broader questions about engaging with the outputs of models that privilege consensus \cite{munnTruthMachinesSynthesizing2023} (i.e., statistically prominent), and dominant narratives (i.e., there is a greater likelihood of bias being reproduced from the underlying training set). Mitigating the challenges of pattern-reproductive functionality of such models is difficult \cite{ferraraEliminatingBiasAI2023} and requires sophisticated and resource-intensive approaches \cite{belroseLEACEPerfectLinear2023}. Addressing LLM bias via techniques such as reinforcement learning also risks \emph{over}-correction; as many commentators have noted, ChatGPT and other models fine-tuned for bias, begin to lose the very expressivity that makes them useful (e.g., \cite{kirkUnderstandingEffectsRLHF2024}). In addition, human feedback on model performance is costly to acquire, while attempts to reduce bias through AI feedback also risks circular results, as the feedback embeds the same bias it aims to monitor \cite{leeRLAIFScalingReinforcement2023, saitoVerbosityBiasPreference2023}.

Even with these impediments, LLMs still offer a complementary analytic power, helping researchers and designers to critically evaluate and engage with material outputs. This paper contributes to thinking about how LLMs can enrich sensemaking and support analytic reasoning, when applied reflexively. Recent work on exploring LLM-enabled TA highlights the importance of creating the `right' prompts \cite{depaoliPerformingInductiveThematic2023, xiaoSupportingQualitativeAnalysis2023} for analysis to be useful, and has emphasised that more work is needed to establish protocols and practices for `prompting' \cite{depaoliPerformingInductiveThematic2023}. Similarly, understanding `how' to use \textit{temperature} \cite{depaoliPerformingInductiveThematic2023} in TA also requires further interrogation, as both convergent (repeatable) and divergent (varied) analytical outcomes can be highly useful in supporting reflexivity and richer analysis. Ultimately, this process should be collaborative: human analysts should not be replaced by AI analysts, but instead work in concert \cite{depaoliPerformingInductiveThematic2023, gaoCollabCoderLowerbarrierRigorous2024, jiangSupportingSerendipityOpportunities2021} - with humans as the ultimate decision-makers.

As highlighted by both this review and related discussions in the literature \cite{depaoliPerformingInductiveThematic2023, zhang2023redefining}, there is rising interest in using LLMs in qualitative research. However, much of this discussion has to date lacked critical consideration of the wider implications of enrolling LLMs into research methods and processes. Though modern LLMs have improved in their handling of overt bias, their `interpretations' may still embed subtle forms of discrimination \cite{mageeIntersectionalBiasCausal2021}. Other work has paid attention to the ``new politics of visual culture'' that arises out of image generation \cite{offert2022sign}, and in a similar vein, indiscriminate LLM use will likely embed unstated political determinations into qualitative research. This is not to deny that qualitative research has always involved ideology. However, the inclusion of generative AI in processes of theme identification and extraction makes it likely that questions around researcher positionality are supplanted by either faith or scepticism in the algorithm's objectivity. This will likely hold implications for trust both in the direct coding analysis itself \cite{depaoliPerformingInductiveThematic2023}, and in the wider circuits of academic production, as AI is increasingly integrated within an `industrialized' automated knowledge industry \cite{van2024big}.

\subsection{Thematic Analysis and Next Token Prediction}\label{thematic-analysis-and-next-token-prediction}

In recent decades, TA has depended on computational support of various text analysis software tools (e.g., \emph{NVivo} \cite{nvivoNVivo2024}, \emph{MAXQDA} \cite{maxqdaAllinoneThematicAnalysis2024}, \emph{ATLAS.ti} \cite{atlas.tiATLASTiAI2024}, \emph{Leximancer} \cite{leximancerLeximancer2024}, \emph{Dedoose} \cite{dedooseDedoose2024}, \emph{Dovetail} \cite{dovetailCustomerInsightsHub2024}, \emph{Voyant Tools} \cite{voyanttoolsVoyantTools2024}). These tools help to manage content, and generate basic thematic outputs, keywords, high-level codes, and word frequencies. With the proliferation of LLMs, new AI-enabled integrations have emerged to support analysis e.g., \emph{ATLAS.ti} developed a suite of AI-enabled tools \cite{atlas.tiAccelerateCustomizeYour2024} built on \emph{OpenAI}'s infrastructure, with features such as: AI-generated code suggestions; intentional AI coding; AI summarisation; and conversational AI, extracting insights by `chatting with your documents'. The volume, size, structure, complexity, and speed at which data is being generated has been a key driver for developing algorithmic approaches to analyse qualitative data. This problem of scale has motivated the desire to improve reliability and validity of qualitative research, harnessing the computational power of LLMs and related language technologies to support effective analysis \cite{gaoCoAIcoderExaminingEffectiveness2023, khanbhaiApplyingNaturalLanguage2021}.

A typical TA process starts with gathering data (e.g., interview transcripts, social media posts, documents etc.) for analysis, often as textual data. The analysts review the data, to familiarise themselves with its content, taking notes and making observations where needed. This is followed by devising and applying a conceptual hierarchy onto the data. This hierarchy can either be pre-configured, developed in advance - using key concepts, or prior literature - or produced more organically - creating the codes while reviewing the data and making adjustments as necessary. Specialist software tools supporting this process enable users to tag discursive markers such as phrases, sentences, and paragraphs with one or more codes. This process is often done over multiple iterations, to recognise and refine patterns and collate them into themes. Such software enables users to then query the collection of documents by theme, or extract individual excerpts, and quotes from the data itself. This enables researchers to assess which themes occur most frequently, and to identify correlated excerpts or quotes. This in turn makes possible what Hackler and Kirsten \cite{hacklerDistantReadingComputational2016} characterise as a mix of both \emph{`distant'} and \emph{`close'} reading of the data.

The training of generative AI has parallels with how we apply thematic analysis to datasets through likely approximations. Prior to the development of \emph{Generative Pre-training Transformer} (GPT) models, made famous by OpenAI's \emph{ChatGPT}, the state-of-the-art models had been `bidirectional', i.e. trained to guess missing data points (e.g., Bidirectional Encoder Representation Transformer (BERT) \cite{devlinBERTPretrainingDeep2019}). Given sufficient training data, time, and scale, models could estimate probability distributions for masked words in ways that approximated human performance (e.g. \cite{devlinBERTPretrainingDeep2019}). This function of estimating a probable missing word is structurally similar to applying a `likely' code, to a fragment of text where no word is missing. This similarity motivated us to explore the possibility of LLMs for suggesting and applying themes to textual data. Prior work (e.g. \cite{depaoliPerformingInductiveThematic2023, mageeAutomatedResearchPractices2023}) explores similar possibilities, but not the specific differences between different language models, nor those between models and human analysts. For this study, we do not use human coders to establish a `ground truth' - our interest is not to evaluate models against human constructed benchmarks. Instead, we focus on exploring the differences that exist between humans and LLM analysis, and reflect upon what these differences might mean in terms of locating and understanding bias.

\section{Methods}\label{methods}

\begin{figure}[ht]
\begin{center}
\centerline{\includegraphics[width=\columnwidth]{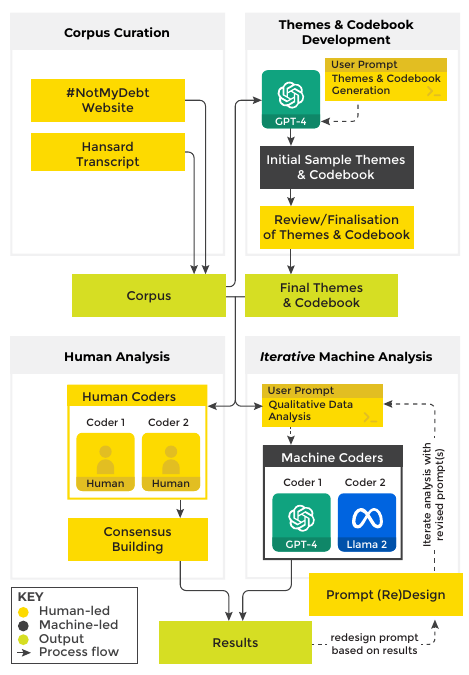}}   
\caption{Overview of Process for LLM-aided Thematic Analysis}
\label{figure1}
\end{center}
\vskip -0.4in
\end{figure}

Our pilot experiment involved collection and analysis of a small sample of statements in relation to the Australian \emph{Robodebt} controversy. These statements were first extracted from social media accounts and government Hansard reports. Using a combination of \emph{GPT-4} and human assessment, a set of 11 themes were then extracted from this sample. The statements were analysed by a pair of human coders (both members of the author team), and by two LLMs, \emph{GPT-4} (June 2023 version) and \emph{Llama 2} (released February 2023). The three sets of results were compared and discussed. Following these results, we re-ran the experiments with revised prompts and models (\emph{GPT-4}, \emph{Claude 3 Opus} and \emph{Llama 3} - all released in April 2024). Figure \ref{figure1} illustrates the major steps in the process.

\subsection{\texorpdfstring{The Case of \emph{Robodebt}}{The Case of Robodebt}}\label{the-case-of-robodebt}

\emph{Robodebt} is the colloquial name for an unlawful automated debt recovery scheme created as an income compliance program by the Australian government, operational from July 2015 to November 2019 \cite{servicesaustraliaInformationRobodebt2023}. This scheme was developed as a replacement to a manual system for calculating welfare over-payments and issuing debt notices to welfare recipients, using an automated data-matching system that compared averaged annual income data from the Australian Taxation Office (ATO) with welfare data from Centrelink \cite{servicesaustraliaInformationRobodebt2023}. \emph{Robodebt} has been criticised by advocacy groups, activists, media, academics, and politicians \cite{CrudeCruelRobodebt2023, ItNotJust, whitefordDebtDesignAnatomy2021}, as well as subject to significant scrutiny through multiple senate committee inquiries \cite{commonwealthAccountabilityJusticeWhy2022}, investigation by the Commonwealth Ombudsman \cite{commonwealthombudsmanAccountabilityActionIdentifying2023, commonwealthombudsmanCentrelinkAutomatedDebt2017}, and a recently completed Royal Commission (Australia's highest form of public inquiry) \cite{commonwealthRoyalCommissionRobodebt2023}. Our rationale for selecting \emph{Robodebt} discourse as the case study was twofold. First, the nature of the discourse was centred around a highly publicised controversy, featuring  differences in political perspectives between those advocating for its legitimacy as an income compliance scheme, and those who protested, investigated and took actions against it. It therefore tests the extent to which LLMs can discern themes that require judgement. Second, although the incident occurred prior to 2021, when data about it might be prevalent, it was not investigated largely until after 2021 - lessening the likelihood of the models having incorporated accounts, excerpts, or training data. The rapid rate of model updates mean there is however no guarantee that later LLM releases do not include these statements in their training data.

\subsection{Data Collection}\label{data-collection}

For the purpose of our experiment, we distilled statements from a corpus of discourse about \emph{Robodebt} from: (1) a transcript of speech by the former Australian Prime Minister Scott Morrison, where he defended his role and position on \emph{Robodebt} in a parliamentary hearing \cite{commonwealthParliamentaryDebatesHansrad2023}; and (2) statements of victims of the Robodebt scheme, who had shared their stories on a volunteer and welfare rights advocate run platform \#NotMyDebt \cite{notmydebtNotMyDebt2018}. Both sources were manually reviewed by two researchers, who then extracted 17 statements based on a text's adherence to certain criteria: (a) the statement needed to present sentiment about the \emph{Robodebt} case, and the broader cultural debate around its impact; (b) the statement had to display a degree of subjectivity and interpretation, ideally value-based rather than a description or statement of fact; and (c) the statement needed to be thematically controversial, thereby likely to evoke divergent views and interpretations. An example statement is as follows (see Appendix \ref{appendix-1-statements} for the full list):

\begin{quote}
Throughout my service in numerous portfolios over almost nine years I enjoyed positive, respectful and professional relationships with Public Service officials at all times, and there is no evidence before the commission to the contrary. While acknowledging the regrettable—again, the regrettable—unintended consequences and impacts of the scheme on individuals and families, I do however completely reject each of the adverse findings against me in the commission's report as unfounded and wrong. \cite[para. 15]{commonwealthParliamentaryDebatesHansrad2023} 
\end{quote}

These criteria provided a diverse group of strongly worded statements that were also feasible for a pilot study involving human coding alongside the evaluation of automated analysis.

\subsection{Developing Themes}\label{developing-themes}

We used a closed thematic codebook for the experiment, which was created using an inductive, hybrid human-LLM approach. Once the corpus had been assembled and read through in depth by the human coders, the corpus was provided to \emph{ChatGPT}, which was then prompted to generate themes based on the guidance for theme generation provided by Braun and Clarke \cite{braunUsingThematicAnalysis2006}. This resulted in the creation of 11 discrete themes, which were then reviewed by the human researchers, to check for contextual validity, with minor semantic adjustments made on one theme. The final 11 themes used for the experiment are presented in Figure \ref{fig2}(A).

\subsection{Constructing the System Prompt}\label{constructing-the-system-prompt}

\begin{figure*}[ht]
\vskip 0.2in
\begin{center}
\centerline{\includegraphics[width=\textwidth]{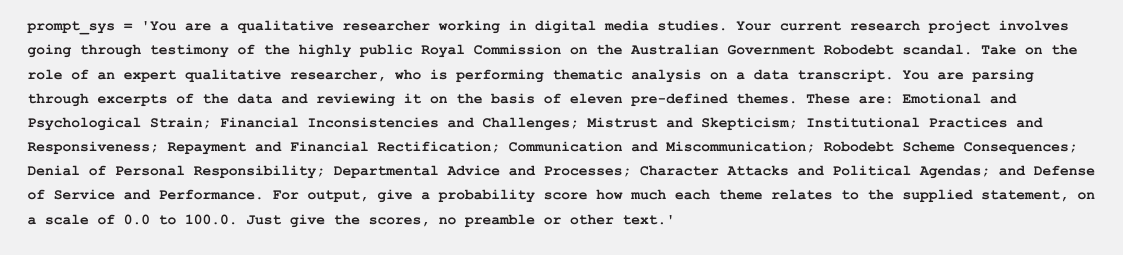}}
\caption{ System prompt provided to the LLMs to conduct thematic analysis \cite{liammageeLLMBiasDetector2023}}
\label{figure2}
\end{center}
\vskip -0.2in
\end{figure*}

The system prompt was created in a series of steps, where we first drafted a set of candidate prompts, and then reviewed, refined, and collated them into a final version. These steps explored phrasing differences that reflected the multidisciplinary composition of the team. For instance, one researcher used minimal technical instructions, using terms like `parse' and requesting `Yes/No' dichotomous responses to theme assignments, while another researcher supplied more contextual information (``\ldots \ involves going through testimonial of the highly public Royal Commission on the Australian Government Robodebt scandal'' ), and asked for scaled ([0––100]) scores for each theme-statement pair. Trial-and-error suggested verbose prompts and quantitative assignments would improve output comparability and interpretability.

Reflections on these draft variations led us to think of this activity as critical to what could be termed `LLM-aided' - an adaptation of the more familiar `computer-aided' - qualitative data analysis approach. \emph{System} and \emph{user} prompts can be considered as artefacts of a collaborative design stage, serving to make explicit questions around researcher commitment, technological constraints, and the context involved in analysis. Rather than a rote formalism, prompt design should be a reflexive and discursive process which translates aspects of the sequential and multi-step nature of thematic analysis \cite{braunUsingThematicAnalysis2006}. Based on our experience, we describe one possible approach to guide this process below (Section \ref{towards-a-collaborative-social-prompt-engineering-ai-sub-zero-bias-cards}).

\subsection{Comparative Analysis}\label{comparative-analysis}

\begin{figure*}[ht]
\vskip 0.2in
\begin{center}
\centerline{\includegraphics[width=\textwidth]{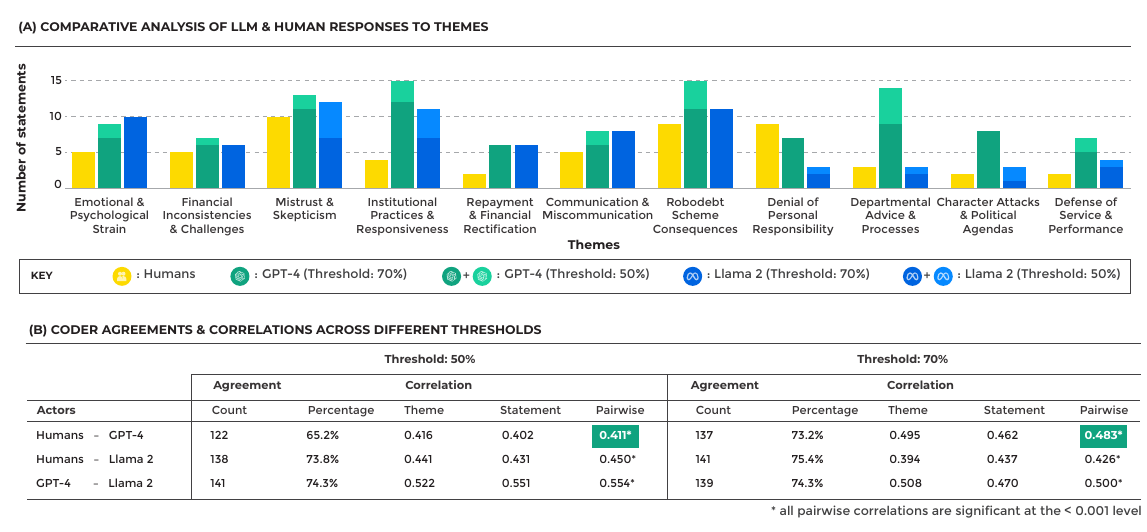}}
\caption{(A) Comparative analysis of LLM and Human Responses to Themes; (B) Comparative Analysis of Coder Agreements, and Correlation of Responses at two different thresholds: 50\% and 70\%.}
\label{figure3}
\end{center}
\vskip -0.2in
\end{figure*}

The \emph{Human Analysis} started with two researchers, independently analysing the statements by indicating the presence or absence of each theme. This was then followed by a collaborative session, where the researchers employed a deliberative, consensus coding method \cite{richardsPracticalGuideCollaborative2018}, systematically reviewing and discussing each statement independently, and arguing for divergent interpretations, with the goal of negotiating through these interpretations to  reach a consensus as to how each statement would be scored. Though time consuming, such consensus coding approaches have been shown to produce reliable coding outputs \cite{richardsPracticalGuideCollaborative2018}.  

For the \emph{Machine Analysis}, we decided to use two widely used LLMs, \emph{GPT-4} and \emph{Llama 2}, opting to use the largest and most recent (as of mid–2023) of each of these models via online APIs. This allowed us to process the prompts and statements in batches, submitting concurrent requests on processing larger volumes of data more efficiently - i.e. we were able to design a prompt that could provide all 17 statements to the LLMs, in addition to the detailed system prompt. The APIs' capacity for customisation and control was also valuable, as we were able to configure the model with parameter values. Another consideration was the reproducibility of results, as the design of the API ensures consistent processing of every request, critical in maintaining the integrity and reliability of the analysis. Accordingly we use a \textit{temperature} setting of 0.1 to minimise variation in both cases. The system instruction prompt \cite{liammageeLLMBiasDetector2023} used for the LLMs is shown in Figure \ref{figure2}.

To \emph{Collate the Analysis}, we took results from all three agents - and developed a result matrix - comprising of the 17 (statements) x 11 (themes) entries. Human entries were dichotomous, indicating presence or absence of a theme, mimicking how coding is often performed in tools like \emph{NVivo}. To assist with interpreting the relative weights according to which LLMs assign themes, machine-generated entries were scored on a continuous scale ([0––100]\%), which was converted into dichotomous values using thresholds. A threshold  of 50\% (See: Figure \ref{fig2}(A)) led to an exaggerated level of theme assignment in both LLM-generated results. We later adjusted this threshold to $\geq$ 70\%, which lowered the number of theme assignments by GPT-4 in particular, and had the effect of increasing the rate of shared thematic assignments and correlations for the three coding agents.

\subsection{Limitations}\label{limitations}

As a pilot study, our results are not intended to be generalisable, but rather could be used to guide other LLM-aided thematic analysis. We note the following limitations and, where relevant, how these limitations might be addressed in future work:

\begin{itemize}
    \item Sample size. Our sample of 17 statements is selective and small, while Thematic Analysis is typically applied to a much larger corpus of text. Larger samples might show different results.
    \item Human coder bias. The qualitative coders identifying the statements in the corpus introduced subjective bias into the corpus identification phase of the research. 
    \item Automated theme extraction. Using GPT itself to extract initial themes from the set of statements may bias results; that particular model is likely to over-identify presence of the same code set in the sample. 
    \item Prompt variations. LLMs are highly susceptible to prompt variations, and alternate prompts may generate different results. We show one example of that below, and note further that, in general, prompts and model parameters can influence model consistency and reproducibility.
    \item The problem of `ground truth'. Though we coded the statements ourselves, as we note below human coding is affected by the same types of bias we are looking to identify. This requires an alternate orientation towards bias, based on consensus-building and triangulation.
    \item Platform evolution. LLMs are new technologies; not only are models released in fast succession, platforms evolve quickly. What appear as model limitations at one point in time - inability to process a certain quantity of text for example - may later disappear, as computational efficiencies, service competition and changing consumer expectations develop. Similarly, statements which are known not to exist in a given model's corpus may later be included as the model is updated.
\end{itemize}

These limitations apply not only to our own study, but also in part to the wider emerging field of automated qualitative research. As we discuss in more speculative terms in our conclusion, this pilot study is in some sense an exercise in teasing out exactly these limitations and constraints.

\section{RESULTS}\label{results}

\subsection{Humans, GPT-4, Llama 2}\label{humans-gpt-llama}

Figure \ref{figure3}(A) shows a comparative depiction of the themes applied per statement from data, across the two different presence probability thresholds: 50\% and 70\% for the LLMs, as compared to human responses. Figure \ref{figure3}(B) shows correlations across both themes and statements are moderate and positive. It also shows total counts of themes applied by statement, again across the same two thresholds. The results indicate that generally there are high levels of agreement across the three agents, with the two machine `analysts' agreeing more often with each other than with the human research group. GPT-4 seemed more likely to apply themes to statements than the other agents. 

Our results also show that asking the machine to score a theme's relevance as a percentile could produce stronger concordance with human analysts, and allows greater transparency and more flexibility in threshold setting. For instance, Figure \ref{figure3}(B) illustrates how GPT-4's agreement with human coders can be increased by raising the threshold value. In practical coding settings, alongside prompts and other parameters this value could be adjusted to increase or decrease agreement, as a measure of reliability or as an indicator of meaningful difference.

At an individual theme level, we also noted that both human and Llama 2 agents were more less likely to apply certain themes, such as \emph{Department Affairs and Processes} and \emph{Character Attacks and Political Agendas}. This could indicate that either GPT-4, in this instance - even with adjusted thresholds - is comparatively greedy in assigning themes; or, conversely, that the human research group was less likely to assign late occurring themes, due to analysis fatigue or a sense that earlier themes had saturated the meaning of certain statements. Different thresholds, prompt variations, and order of statements / themes could work to increase inter-coder reliability and improve confidence in the overall thematic analysis.

We also observed variances in content coding tendencies between the human and machine analysis. Certain themes, such as \emph{Mistrust and Skepticism}, and \emph{Robodebt Scheme Consequences,} showed remarkable agreement between the machine and humans. In the case of others like \emph{Denial of Personal Responsibility} - where \emph{Llama 2} barely assigns the theme - we observed significant discrepancy. It is possible this difference results from the need for an additional step, to read against the `grain' of written text - the product of an interpretive suspicion that draws upon lived experiences. Unless explicitly instructed to do so (see \ref{gpt-4-revisited}), the LLMs appear to be less likely able to identify the latent meaning, where a second-order interpretation (`I think this speaker is actually denying that they are personally responsible') is required. In contrast, both LLMs were able to ascribe negatively worded explicit themes such as \emph{Mistrust and Skepticism} more clearly. This variance may also expose underlying subjective biases held by the human researchers. In contexts like this, the machine responses could elicit further reflection of the interpretations of the human analysts. Finding ways to explore this through cautious adversarial prompting is an important area for future work.

\subsection{The Reflexive LLM: `Be Sceptical', `Be Parsimonious' and `Take your time'}\label{gpt-4-revisited}

Intensive LLM development and experiment with prompt design has resulted in support for longer text prompts step-by-step reasoning. In April 2024, we re-ran our comparison with the most recent release (April 9, 2024) of \emph{GPT-4}, an updated version of \emph{Llama3} (70 billion parameter model) and \emph{Claude 3}. The previous results had shown LLM agents were more likely to assign themes to statements compared to human coders, and were also less likely to assign themes that involved a sceptical interpretation of a statement, such as \emph{Denial of Personal Responsibility}. Accordingly we added two instructions: `Be sceptical' and `Be parsimonious'. We also requested each model to generate the full table of theme-to-statement results two times; on the second iteration, we asked it to `Take your time' to reflect upon and, if necessary, revise its initial scores. We included an instruction to justify why scores had been modified.

\begin{figure*}[ht]
\vskip 0.2in
\begin{center}
\centerline{\includegraphics[width=\textwidth]{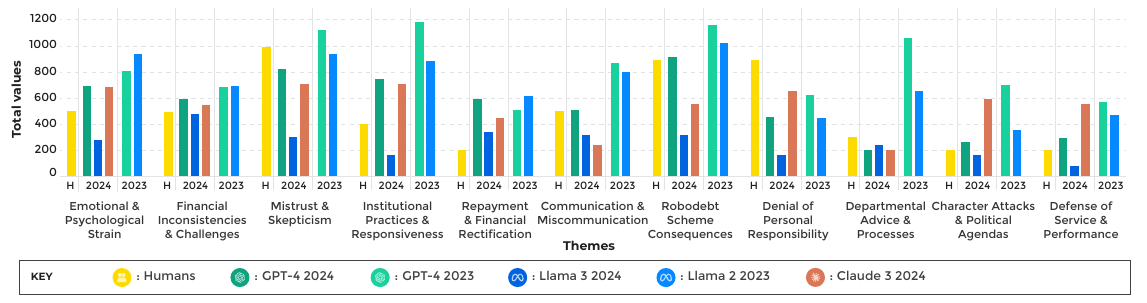}}
\caption{Comparison, with 2024 models and modified instructions}
\label{figure4}
\end{center}
\vskip -0.2in
\end{figure*}

\begin{figure*}[ht]
\vskip 0.2in
\begin{center}
\centerline{\includegraphics[width=\textwidth]{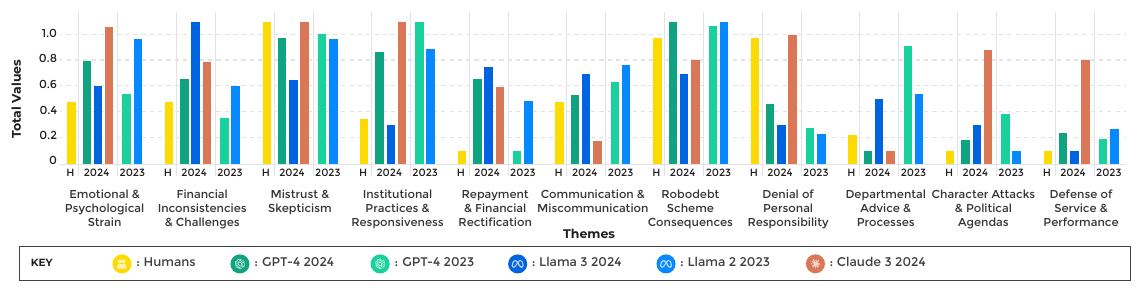}}
\caption{Comparison, with 2024 models and modified instructions - scores standardised}
\label{figure5}
\end{center}
\vskip -0.2in
\end{figure*}

Figures \ref{figure4} and \ref{figure5} show results with the three 2024 models added – with Figure \ref{figure5} using standardised scores to indicate relative frequency. Similar to the human coders, the new model/prompt combinations are less likely to assign themes to statements. In certain cases – \emph{Denial of Personal Responsibility}, \emph{Communication and Miscommunication}, \emph{Defence of Service and Performance} – either \emph{GPT-4} (2024) or \emph{Claude 3} assigned themes with similar frequency to the human coders, in both absolute and relative terms. Tests showed high positive correlations ($> 0.5$, $p \textless 0.01$) for the same two models; curiously, \emph{Llama 3} had correlated more weakly than any of the other models, including \emph{LLama 2}.

The generated reasons for score revision suggested how the influence of observed differences between human and machine coding can lead to refined prompting that, in effect, helps to align model with human outputs. Below are two examples:
 
    \begin{quote}
        \textbf{\emph{(GPT-4)}}: Reduced scores in less relevant themes. Scores were reduced or removed in themes that were less central to the content of the statements to focus on the most impactful themes. 
    \end{quote}

    \begin{quote}
        \textbf{\emph{(Claude 3)}}: For statements that appeared to be couched in rhetoric or excuse-making, I have increased the scores for themes like `Denial of Personal Responsibility’ and `Character Attacks and Political Agendas’. These statements tended to downplay personal involvement in the Robodebt scheme and shift blame to others. 
    \end{quote}

As we discuss below, score differences can also be used to further calibrate human coding.

\section{DISCUSSION}\label{discussion}

\subsection{\texorpdfstring{Empirical Outcomes \& Methodological Reflections }{Empirical Outcomes \& Methodological Reflections }}\label{empirical-outcomes-methodological-reflections}

These pilot findings hold implications for the future role of LLMs in supporting thematic analysis. In the main experiment, both models were able to generate and apply themes to textual content meaningfully, in ways that can `prompt' human qualitative researchers about their own judgements. In these results, the models were also more closely aligned with each other than with human coders. Applying an updated \emph{GPT-4} model and modifying the initial prompt produced results more inline with human coding, when scores are aggregated by theme - though individual assignments actually varied more. This suggests that iterated prompting can align LLM results more closely while also raising new questions: were the human coders too eager to identify statements as reflecting \emph{`Denial of Personal Responsibility'} for example? Or is this difference a result of the LLM being reluctant to judge statements negatively? And if so, can this response be `nudged' through additional prompting? 

In addition, we note that presence of prior information on a specific topic like \emph{Robodebt} may distort findings - models may for instance draw upon other references in ways that perturb qualitative assessment. Further work is needed on what effects, if any, this produces, compared to a `novel' topic of which the model has no prior knowledge. Using themes suggested by the model itself indicates that there is some notion of what is `known' about the data, which although sensible for this particular experiment, could potentially occlude interesting and salient points from the data that might not neatly fit into the predetermined categories - in the present experiment there is no provision for more constructivist approaches to engaging with the data. As human analysts' role in the initial identification of codes was limited to sense-checking for relevance - and not identifying the codes themselves - further work is required to test how LLMs perform on `unseen' codes. Use of \emph{Llama2} acts as a control to some degree, but it is also possible, given how models are increasingly trained on the output of other models, the biases of \emph{GPT-4} carry over. Other tests could be used; for example, human researchers could review a subset of the corpus to create initial themes, and then use LLMs to apply them. Such tests might locate blind spots and biases. 

Often however this approach may be incompatible with the ill-defined, wicked problems human-centred research often deals with, where variables are unknown and interpretation involves reflexivity  \cite{rittelDilemmasGeneralTheory1973}. The second experiment suggested that models can be `coaxed' into greater alignment with human coding, and while the language of those refined prompts was generic, we noted this still involves several iterations to ensure prompt instructions are followed accurately. Paradoxically, automation may involve more rather than less time and labour: effort is required to identify or review codes, for example, to ensure they are salient and appropriate. `Fact-checking' both theme identification and assignment grows dramatically with corpus size, and evaluation becomes a second-order order problem: machine `coding' requires cross-checking against human results, and this may be prohibitive in many research settings. Although there is guidance for conducting TA, there is also no single `correct' way of interpreting text or other data; and differences between machine and human interpretations may be confounding rather than clarifying. In other contexts, LLMs may be useful as possibility engines, showing diverse perspectives or acting as Socratic testers of human assumptions. As we note above, they have potential for encouraging divergent \cite{banathyDesigningSocialSystems1996} or designerly ways of thinking \cite{crossDesignerlyWaysKnowing1982}, and can help researchers to reflexively engage with different arguments \cite{rittelSecondGenerationDesignMethods1984}, learn from the material `back-talk' \cite{schonReflectivePractitionerHow1983}, and use abductive reasoning \cite{kolkoAbductiveThinkingSensemaking2010}. In our pilot, differences in how LLMs coded themes did provoke questions about how humans coded them, and we anticipate this would remain true in larger scale, iterated thematic analysis. For example, differences in human and machine applications of the theme \emph{Denial of Personal Responsibility} prompted reflection on why human coders were more likely to apply this theme. In hindsight, we considered human judgement as more accurate in this instance, and as the second experiment showed, simple adjustments to the prompt (``Be sceptical'') were enough to align outputs with those judgements. In the case of a more descriptive theme (\emph{Repayment and Financial Rectification}), all of the models in both experiments were more likely to assign it; and in this instance we agreed with the machinic interpretation. Here a bias - towards finding deeper or more evocative interpretations - might mean humans are less inclined to assign apparently boring themes. Use of LLMs has potential to widen the epistemic position of researchers, forcing them to confront their own biases.  

A further consideration relates to replication. Inclusion of LLMs amplifies the complexity, or degrees of freedom, involved in the TA situation. Not only do human researchers differ in how they code; different LLMs also vary and, as we have noted, are subject to a host of technical conditions, including the precise wording and sequencing of prompts, as well as model parameters such as the temperature setting (with higher settings increasing randomness and lowering output reproducibility). This complexity may have the perverse effect of reducing the credibility of computer-aided qualitative research, as it becomes, in either perception or actuality, a hybrid of human-machine interpretation. To preempt such understandable lines of criticism, we view it as important to integrate LLMs as invariably flawed complementary aids rather than as newly-found infallible oracles or truth-tellers that could replace human interpretation.

Though conducted over a small dataset, our results underscore the potential for applying the same technique on larger corpora. LLMs appear to offer promise in support of TA and other common forms of computer-aided qualitative data analysis, especially when the corpus is large or, as is increasingly the case, project budgets are constrained. However we view use of LLMs as requiring caution, with researchers being cognizant that over-reliance on such tools can limit reflexivity and skill development \cite{braunOneSizeFits2021}, as well as restrict their authority and control over the analytic process \cite{nowellThematicAnalysisStriving2017}. While subjective human reasoning is itself a black-box, human researchers are accustomed to supporting (or at least explaining) their reasoning with evidence. Conversely, the black-box nature of LLMs makes it difficult to understand how or why specific codes and themes are assigned, and this difficulty only grows with model size and complexity \cite{carabantesBlackboxArtificialIntelligence2020}. This suggests a need to develop new skills to prompt and interrogate these models, in order to understand how they generate their responses. As the second experiment suggests, prompt variations - including requests to ``take your time'' or ``go step-by-step'', employing chain-of-thought and other reasoning techniques - can supply at least heuristic guidance as to how a model arrives at certain outcomes. However, further developments in explainable AI, coupled with human oversight and intervention, will be crucial to ensuring such automated techniques are not viewed with the suspicion they understandably deserve today.

\subsection{\texorpdfstring{Socialising Prompt Engineering - `Sub Zero' Bias Cards }{Towards a Collaborative, Social Prompt Engineering - AI Sub Zero Bias Cards }}\label{towards-a-collaborative-social-prompt-engineering-ai-sub-zero-bias-cards}

\begin{figure*}[ht]
\vskip 0.2in
\begin{center}
\centerline{\includegraphics[width=\textwidth]{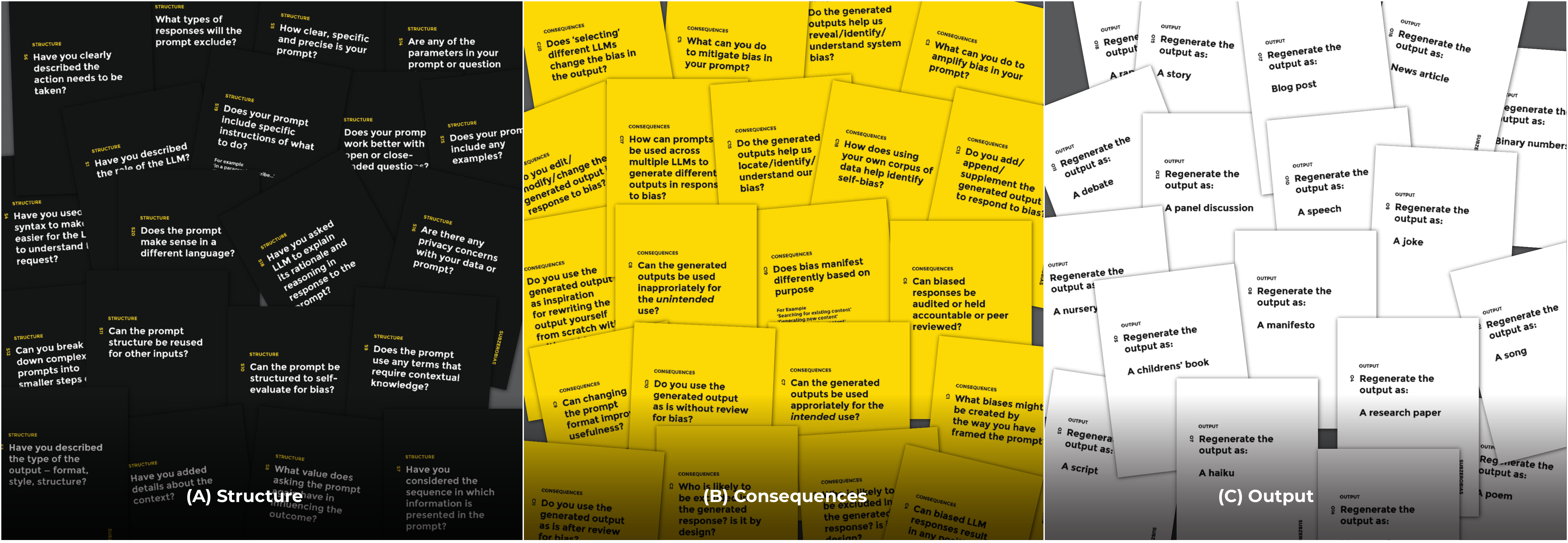}}
\caption{AI Sub Zero Bias Cards}
\label{fig6}
\end{center}
\vskip -0.2in
\end{figure*}

The mature world of computer-aided qualitative data analysis software (CAQDAS) has, unsurprisingly, begun to integrate LLMs into products (e.g. \cite{atlas.tiATLASTiAI2024}), and the promise of efficiency makes further progress seem inevitable. Our findings echo those of other researchers \cite{depaoliPerformingInductiveThematic2023, xiaoSupportingQualitativeAnalysis2023, gaoCollabCoderLowerbarrierRigorous2024, jiangSupportingSerendipityOpportunities2021} in highlighting the importance of including `humans-in-the-loop' when using LLMs as analytic tools. We also underscore the value of reflexivity and collaboration during prompt design and evaluation of theme assignment. 

While De Paoli \cite{depaoliPerformingInductiveThematic2023} has offered general recommendations, the existing literature provides little concrete guidance on what researchers need to do in these adaptations of traditional thematic analysis. In this section we outline an approach to support prompt design and evaluation. Termed the \href{https://espace.library.uq.edu.au/view/UQ:f998cc0}{\emph{AI Sub Zero Bias Cards}} \cite{khanAISubZero2023}, the approach uses questions placed on cards to guide, provoke and `prompt' human researchers during these phases. As well as leaning upon the extensive literature on algorithmic bias and the political entailments that follow AI use \cite{angwinMachineBias2016, saitoVerbosityBiasPreference2023, algorithmicjusticeleagueMissionTeamStory2021, mageeIntersectionalBiasCausal2021, offert2022sign, Amoore_2023}, these questions follow from our own practice and experimentation.

Our own process of prompt design involved many iterations, with differences in syntax, structure and perspective, producing different results. As we noted above, initially LLMs appeared less likely to `read into' statements, instead taking in particular official statements at face value. This in turn led us to include instructions to be `sceptical', to align model outputs more closely with human coders. This in turn perturbs the models towards other - more explicit - forms of bias, and we had to question our own assumptions in doing so. The card toolkit asks researchers to consider the effects of such suggestions in a reflexive way, and in doing so, differs from the types of guidance offered by LLM vendors and designers (for example, \emph{OpenAI} recently authored their own prompt engineering user guide \cite{openaiPromptEngineeringGuide2023}).  We also wanted to operationalise these concepts to encourage exploration and discussion, and hence decided to create a generative card-based toolkit to facilitate the process.

The toolkit comprises 58 provocations, distributed across three card types: (1) \emph{Structure}; (2) \emph{Consequences}; and (3) \emph{Output}. The \emph{Structure} cards interrogate the decisions made during the prompt construction, i.e., the semantic phrasing, organisation, and configuration of the prompt itself, which can be consequential to the simulation of `perspective' embedded in the thematic analysis. The \emph{Consequences} cards provoke the user to reflect on the outcomes generated by the prompts and their implications. The \emph{Output} cards draw on the principles of creativity such as random stimuli \cite{debonoSeriousCreativityUsing1992}, suggesting experimentation with restructuring and reformatting the prompt outputs in unconventional forms to foster a sense of play and reflexivity through the dialogue \cite{brandtFormattingDesignDialogues2008}. These help to develop a `liminal' space for reflection \cite{turnerLiminalityCommunitas1969}, as well as engaging researchers more deeply with the content when viewed through an unfamiliar frame \cite{haraDesigningDesign2018}. 

The card-based toolkit offers a distillation of our own experience, and can be used flexibly, to create modular, dynamic and accessible frames for complex interpretive tasks and processes \cite{hsiehWhatCardsExploring2023, royCardbasedDesignTools2019}. For instance, it can enable researchers to experiment with prompts to better appreciate how biases can emerge through LLM engagement, helping to address issues of AI explainabiltiy and transparency. Similar to how Hornecker \cite{horneckerCreativeIdeaExploration2010} has described the utility of game cards, such systems can be used as creative avenues for experimentation and analysis. The \emph{AI Sub Zero Bias Cards} can also be used as participatory tools \cite{sandersFrameworkOrganizingTools2010}, formatting collaborative design games and dialogue \cite{brandtFacilitatingCollaborationDesign2004, brandtFormattingDesignDialogues2008} to facilitate interactions between researchers and LLM-based agents \cite{atlas.tiATLASTiAI2024}. As LLMs become integrated into qualitative analysis, materials like these can serve as educational and exploratory devices, but can also guide more systemic investigation and `red teaming'\footnote{Adversarial role playing to identify vulnerabilities or weaknesses of systems, to improve blind spots or gaps.}, to identify opportunities for LLM alignment and bias mitigation.

\section{Conclusion}\label{discussion-conclusion}

Our paper makes several contributions. First, it adds to growing attention to how LLMs can be applied to thematic and other forms of qualitative analysis  \cite{depaoliPerformingInductiveThematic2023, gaoCollabCoderLowerbarrierRigorous2024, jiangSupportingSerendipityOpportunities2021}. In examining a controversial topic that has received considerable Australian media coverage, it identifies differences between how humans and LLMs engage critically with textual data. Alongside the fast-changing nature of LLM development, the second experiment also illustrates the sensitivity of LLMs to prompt variations. Compared to other algorithmic approaches, this sensitivity means researchers may need to spend more time experimenting with options - different models, parameters and prompts - to understand their often subtle effects. We also note that asking LLMs to output scaled values permits different thresholds, which can modulate levels of agreement between human and machine coders, in ways that could prove helpful in either underscoring agreements or highlighting differences. Similar to our pilot study, for large-scale LLM-aided thematic analysis we suggest conducting pilots that iterate through and evaluate prompt and model combinations with a small sample. 

The \emph{AI Sub Zero Bias} card toolkit collates our reflections and responds to previous work highlighting the importance of refining prompts for TA. Further work is needed to evaluate how such tools can support prompt design and analysis of outcomes in the context of LLM-aided thematic analyses. Beyond assessment of individual approaches, this may also illuminate how the very act of interpretation is changing through a human-machine hybrid collaborative analysis \cite{depaoliPerformingInductiveThematic2023, gaoCollabCoderLowerbarrierRigorous2024, jiangSupportingSerendipityOpportunities2021}. Such evaluations could include: larger corpora and more intensive cases of thematic analysis; a wider range of models, parameters, prompts and threshold settings (when scaled outputs are requested); and different examples of topics and themes. 

Finally, thematic analysis is an iterative, multi-stage process, and this study focuses on a specific aspect of this process. Rather than attempting to develop a valid method which entrusts QDA entirely to machines, this study lays groundwork for researchers to critique and understand their own epistemic assumptions and commitments. Consequently, it contributes to the broader conversation regarding the role of technology and human subjectivity in qualitative research.

\subsection{\texorpdfstring{Towards `Thematic Agents'?}{Towards `Thematic Agents'?}}\label{towards-gpt-agents}

We conclude with a brief discussion of emergent tendencies in language models, and consider several of their wider implications. In late 2023 OpenAI released \emph{GPTStore}, a marketplace for customisable LLM-based `agents' \cite{logankilpatrickGPTBuilder2024, natalieCreatingGPTOpenAI2024} that signalled one possible direction for how LLMs can be adapted to specific tasks. While `GPTs' featured on \emph{GPTStore} involve little more than extended prompts, they illustrate how language model behaviour might be tailored to tasks and user preferences. The addition of a `memory' feature to \emph{ChatGPT} in 2024 also suggests how LLMs might retain a sense of these preferences across individual chat interactions. While OpenAI is only one vendor, in the context of thematic analysis these extensions show how LLMs can be integrated into platforms that align with individual or community preferences. 

On the one hand, this tendency toward personalisation means that LLM outputs can be progressively adapted to unique researcher practices and support development of what Braun and Clarke term  `analytic (craft) skills' \cite{braunOneSizeFits2021}. Contrary to what is sometimes misread as a `prescriptive' phased approach to thematic analysis, this personalisation could encourage researchers to develop a unique interpretative praxis, following Braun and Clarke's advice to imagine analysis as a heuristic and iterative exercise: ``as one's analytic (craft) skill develops, these phases can blend together somewhat, and the analytic process necessarily becomes increasingly recursive'' \cite{braunOneSizeFits2021}. For example, a thematic agent could build upon a history of coding practice as it introduces researchers to new data sets and issues, reviews and queries theme human assignments, or modifies its biases when suggesting its own candidate themes. Responses to the provocations introduced in the card toolkit could also be embedded as `preferences' that condition the LLM's future behaviour. 

On the other hand, LLMs - especially when hosted, as is the case with the most powerful models, on remote commercial infrastructure - introduce new ethical and political questions. The interplay between model training, reinforcement learning, prompt wording and the data set used in thematic analysis is complex, and can lead to biased results whose causes are difficult or impossible to isolate. Thematic analysis often involves sensitive data that should not be transmitted to third-party services, and some processes of de-identification (e.g. paraphrasing) may also modify LLM outputs in subtle ways. At a larger scale, and despite the recommendations of scholars to retain `humans-in-the-loop', LLM performance may convince funders of research to bypass or shortcut human review. The automated `copilot' may in other words take full control of the interpretative vehicle. As Amoore \cite{Amoore_2023} has noted, reliance upon LLM and other automated systems can lead to an often surreptitious modulation of social, political and epistemic norms. Thematic analysis is one of many contexts where this modulation must itself be modulated by ongoing human scrutiny and review.

As a consequence, we stress that this `agency' of LLMs should be regarded as an opportunity for the provocation – rather than delegation – of interpretation. Our findings elaborate on existing work suggesting these models hold potential as powerful reflexive instruments   \cite{schonReflectivePractitionerHow1983} to scaffold and augment human capabilities for sensemaking and abductive reasoning \cite{kolkoAbductiveThinkingSensemaking2010} in thematic analysis. While it is likely these tools will play roles in future humanities and social science research, the critical faculties developed in these disciplines also need to be brought to bear upon the technologies themselves. Approaches like \emph{AI Sub Zero Bias Cards} serve as pragmatic spurs for an ongoing mutual interrogation between human and machinic researchers. Further work is required to explore alternate pathways for pursuing this interrogation.

\section{Acknowledgements}

This research was conducted by the ARC Centre of Excellence for Automated Decision-Making \& Society (CE200100005) and  funded by the Australian Government through the Australian Research Council. We would like to acknowledge especially the continued support of Sally Storey and other colleagues at the ADM+S. We would also like to thank the anonymous ACM Conference on Human Factors in Computing Systems (CHI) and Microsoft Journal of Applied Research (MSJAR) reviewers for their constructive feedback and suggestions, which have helped to improve the manuscript's clarity and quality.

\bibliographystyle{ieeetr}
\bibliography{MSJAR}

\appendix
\section{Robodebt Statements}\label{appendix-1-statements}

Statements are taken from Australian Hansard record and \url{https://www.notmydebt.com.au/}.

\begin{enumerate}
\item After I cancelled my payment they paid me extra money, I was actually entitled to it but they tried to say it was a debt they also tried to pay me money I was not entitled to and refused to stop the payment (even though I was asking them to stop the payment before it happened).
Copy code\item Centrelink contacted me in 2018 claiming I owed \$1950 due to misreporting my income while on Newstart during the 2014/15 financial year. I disputed the debt but lost so had to repay the full amount. Centrelink has sent me a letter today stating that: ``We are refunding money to people who made repayments to eligible income compliance debts. Our records indicate that you previously had debt/s raised using averaging of ATO information. We no longer do this and will refund the repayments you made to your nominated bank account.'' Hell yes!

\item Throughout my service in numerous portfolios over almost nine years I enjoyed positive, respectful and professional relationships with Public Service officials at all times, and there is no evidence before the commission to the contrary. While acknowledging the regrettable—again, the regrettable—unintended consequences and impacts of the scheme on individuals and families, I do however completely reject each of the adverse findings against me in the commission's report as unfounded and wrong.

\item The recent report of the Holmes royal commission highlights the many unintended consequences of the robodebt scheme and the regrettable impact the operations of the scheme had on individuals and their families, and I once again acknowledge and express my deep regret for the impacts of these unintended consequences on these individuals and their families. I do, however, completely reject the commission's adverse findings in the published report regarding my own role as Minister for Social Services between December 2014 and September 2015 as disproportionate, wrong, unsubstantiated and contradicted by clear evidence presented to the commission.

\item As Minister for Social Services I played no role and had no responsibility in the operation or administration of the robodebt scheme. The scheme had not commenced operations when I served in the portfolio, let alone in December 2016 and January 2017, when the commission reported the unintended impacts of the scheme first became apparent. This was more than 12 months after I had left the portfolio.

\item The commission's suggestion that it is reasonable that I would have or should have formed a contrary view to this at the time is not credible or reasonable. Such views were not being expressed by senior and experienced officials. In fact, they were advising the opposite.

\item At the last election, Labor claimed they could do a better job, yet Australians are now worse off, paying more for everything and earning less—the exact opposite of what Labor proposed. For my part, I will continue to defend my service and our government's record with dignity and an appreciation of the strong support I continue to receive from my colleagues, from so many Australians since the election and especially in my local electorate of Cook, of which I am pleased to continue to serve.

\item Media reporting and commentary following the release of the commission's report, especially by government ministers, have falsely and disproportionately assigned an overwhelming responsibility for the conduct and operations of the robodebt scheme to my role as Minister for Social Services. This was simply not the case.

\item Over \$20,000 debt dating back to 2012. In that time I was working casual, doing courses and also homeless. I had 2 children to worry about. All my tax returns where taken from me and any FTB. I had a breakdown in 2016. I have lived with stress since the start of all the debts coming in, 9 in total!

\item I was hit twice by the RoboDebt scheme. The first year they stated I owed money from an employment role in 2008. I was working as a Cadet getting Study Allowance alongside my Salary — Centrelink calculated that I earned \$8000 in 8 weeks. What a laugh! I am a single parent who could only dream of earning that kind of money. They sent me a debt letter of \$3600. I have paid that despite the fact that I knew I did not owe it, I did not want the stress and anxiety — just working to make ends meet as it is. 

\item I am a single mum and due to this debt have a record so it's hard to find work now. I live in a private rental home and do not work so it is very stressful. I have paid money to it since 2014 and all lump sums in July have gone on it.

\item I kept getting phone calls, a number i didn't recognise, 3-4 times a week. When i answered it would be prerecorded message, an American accent telling me I needed to contact some legal firm, when I called the number, i'd get another pre-recorded message.

\item I broke both my legs and was in a wheelchair for months and I work as a chef I had to prove I wasn't working, and told me that I declared that I made \$0 that year which is a lie gave me \$5500 debt I asked for evidence several time with no success. Might I add I've work all my adult life first time I really need centerlink then I worked my arse off to be able to walk again and earn my money just to get back to work.

\item I also noted in evidence departmental statistics on the sole use of income averaging to raise debts under Labor ministers Plibersek and Bowen and form and actual letters used by the department going back as far as 1994 that highlighted this practice. The evidence I provided to the commission was entirely truthful.

\item Robodebt has shaken not only my trust but the trust of our society in the Australian Public Service. I know that the frontline workers do their best, in sometimes very difficult circumstances, to deal with the public who are very stressed, but there was a complete failure of leadership in the higher echelons of the Public Service and a complete failure of political courage and political understanding of the importance of providing support to the most disadvantaged in our society. 

\item I am still shocked by the response of the previous government, and I still cannot understand why they pushed forward over a number of years in this process. Despite any advice about how bad the Centrelink retrieval of debt process was, they still refused to act, and they should hang their heads in shame about it.

\item In 2021, I spoke in this place about how my electorate of Macarthur had lost people to suicide because of the stress that robodebt had placed upon them. I saw it firsthand. People in my electorate felt and lived firsthand how the former coalition government and those senior public servants who backed in this terrible scheme did not care for them, their families or their attempts to deal with such a pathetic witch-hunt, known as robodebt.
\end{enumerate}

\end{document}